\documentclass[aps,superscriptaddress,showpacs]{revtex4}
\usepackage{amsmath}

\begin{document}
\bibliographystyle{apsrev}

\title{Sensitivity optimization in quantum parameter estimation}

\author{F. Verstraete}
\affiliation{Institute for Quantum Information, California
Institute of Technology, Pasadena, CA 91125}
\affiliation{SISTA/ESAT, Katholieke
Universiteit Leuven, K. Mercierlaan 94, Leuven, Belgium}
\author{A. C. Doherty}
\author{H. Mabuchi}
\affiliation{Institute for Quantum Information, California
Institute of Technology, Pasadena, CA 91125}
\date{\today}

\begin{abstract}
We present a general framework for sensitivity optimization in quantum
parameter estimation schemes based on continuous (indirect) observation of a
dynamical system. As an illustrative example, we analyze the canonical
scenario of monitoring the position of a free mass or harmonic oscillator to
detect weak classical forces. We show that our framework allows the
consideration of {\em sensitivity scheduling} as well as estimation
strategies for non-stationary signals, leading us to propose corresponding
generalizations of the Standard Quantum Limit for force detection.
\end{abstract}

\maketitle

The primary motivation for work presented in this paper has been to
contribute to the continuing integration of quantum measurement theory with
traditional (engineering) disciplines of measurement and control.  Various
researchers engaged in this endeavor have found that the concepts and methods
of theoretical engineering provide a fresh perspective on how differences and
relationships between quantum and classical metrology can be most cleanly
understood.  This approach has been especially fruitful in scenarios
involving {\em continuous} measurement, for which a number of important
physical insights and results of practical utility follow simply from the
formal connections between quantum trajectory theory and Kalman filtering
\cite{wiseman1995,Milburn96,mabuchi1996b,mabuchi1,Doherty1,Doherty99c,Belavkin99}.

Here we describe a general formalism for parameter estimation via continuous
quantum measurement, whose equations are amenable to analytic and numerical
optimization strategies.  In addition to being useful for practical design of
quantum measurements, we find that this approach sharpens our understanding
of the significance and origin of Standard Quantum Limits (SQL's) in
precision metrology. Following the basic notion that the ``standard limit''
for any measurement scenario should be derivable by optimization over some
parametric family of ``standard'' measurement strategies, we present results
that generalize the SQL for force estimation through continuous monitoring of
the position of a test mass. Our analysis shows that the canonical expression
for the force SQL in continuous position measurement stems from a rather
arbitrary limitation of the set of allowable measurement strategies to those
with constant sensitivity, and we find that a lower expression (by a factor
of $3/4$) can be obtained when time variations are allowed. It follows that
further expansions of the optimization space (such as adaptive measurements
with real-time feedback \cite{wiseman1995}) should be considered in order to
arrive at an SQL that consistently accounts for a natural set of measurement
strategies that are ``practically equivalent'' in terms of inherent
experimental difficulty. 

For clarity, the main results of this paper are presented in the first and
third sections within the concrete context of force estimation via continuous
position measurement. In order to emphasize the general nature of our
formalism and the conclusions we derive from it, the second section provides
a more abstract development that arrives at all the equations needed for
sensitivity optimization in a broad class of continuous measurement
scenarios. As this general treatment is rather technical, we note that it is
not crucial to the overall logical flow of the paper. Very recently
Gambetta and Wiseman have 
discussed a similar approach to
parameter estimation for resonance fluorescence of a two-level atom
paying particular attention to how information about the unknown
parameter, and also about the quantum state, changes with different
kinds of measurements~\cite{gambetta2001a}. 

\section{Force estimation by continuous measurement of position}
The aim of this section is to present a formalism for continuous parameter
estimation in the specific context of a harmonic oscillator subject to an
unknown force linear in $\hat{x}$. This section gives a rigorous and a more
general treatment of the ideas previously worked out by one of us
\cite{mabuchi1}. We first derive the conditional evolution equations for the
oscillator under continuous position measurement, then discuss their
control-theoretic interpretation as Kalman filtering equations. We then show
how a Bayesian parameter estimator can be obtained from the Kalman filter in
this scenario.

\subsection{Conditional evolution equations}
We will derive the equations of motion of a continuously observed system
conditioned on the measurement record. Our treatment is based on the model of
continuous measurement of Caves and Milburn \cite{caves1987a}, which in turn
was based on work of Barchielli {\it et al} \cite{Barchielli}. Their
derivation is solely based on the standard techniques of operations and
effects in quantum mechanics which makes it very transparent.  Similar
results could have been obtained by making use of the quantum-stochastic
calculus of Hudson \cite{Hudson} as was done by Belavkin and Staszewski
\cite{Belavkin}.

In continuous measurement --- often an accurate description of
experimentally realizable measurements ---  projective collapse of the
wavefunction, and hence also the Zeno effect, can be
avoided by continually performing infinitesimally weak measurements. A weak
measurement consists of  weakly coupling the system under interest to a
(quantum-mechanical) meter, followed by a von Neumann measurement of the
meter state. As there was only a weak coupling, only very little information
about the system of interest is revealed and there will only be a limited
amount of back-action. At first we will introduce the concept of weak
measurements in the framework of position measurement. Then we will show how
to derive the equations of motion for a quantum particle subject to a whole
series of weak measurements. The treatment of continuous measurements will
then be obtained by taking appropriate limits.

The aim of a weak position measurement is to get some information out of the
system, although without disturbing it too much. This can be done by applying
a selective POVM $\{\hat{A}_{\xi}(\hat{x})\}$ where there is a
lot of overlap between the $\hat{A}_{\xi}(\hat{x})$ associated with different
measurement results $\xi$. This overlap is proportional to the variance of
the measurement outcome, but inversely proportional to the variance of the
back-action noise. As shown by Braginsky and Khalili \cite{Braginsky}, the
product of those variances always exceeds $\hbar^2/4$. Equality is achieved
if and only if $\hat{A}_\xi(\hat{x})$ is Gaussian in $\hat{x}$. As we
are interested in 
the ultimate limits imposed by quantum mechanics, we will assume our
measurement device is optimally constructed so as to yield a Gaussian
$\hat{A}_\xi(\hat{x})$:
\[\hat{A}_\xi(\hat{x})=\frac{1}{(\pi
D)^{1/4}}\exp\left(-\frac{(\xi-\hat{x})^2}{2D}\right)\] This is equivalent to
the model of Barchielli and also of Caves and Milburn
\cite{caves1987a} who obtained it by explicitly working out the case
of linear
coupling between a (Gaussian) meter and the particle followed by a von
Neumann measurement on the meter.

We will now assume that the wavefunction of the observed particle is also
Gaussian. This is a reasonable assumption as we will soon take the limit of
many Gaussian measurements, each of which effects a Gaussian ``conditioning''
of the particle's wavefunction. Ultimately the wavefunction itself will
become Gaussian, whatever its original shape. We furthermore assume that the
Hamiltonian of the unobserved particle would be given by:
\begin{equation}H_0=\frac{\hat{p}^2}{2m}+\frac{m\omega^2}{2}\hat{x}^2+\theta
\hat{x},
\end{equation}
where $\theta$ is the (eventually time-dependent) force to be
estimated. It will turn out to be very useful to parameterize the
Gaussian wavefunction of the particle by a complex mean
$\tilde{x}=\tilde{x}_r+i\tilde{x}_i$ and complex variance
$\tilde{\sigma}=\tilde{\sigma}_r+i\tilde{\sigma}_i$ (throughout
the paper the notation $\sigma$ instead of $\sigma^2$ will be used
to denote the variance):
\begin{eqnarray}|\psi\rangle&=&|\tilde{x}(t),\tilde{\sigma}(t)\rangle\nonumber\\
\langle x|\psi\rangle&=&\left(\frac{\tilde{\sigma}_r}{\pi|\tilde{\sigma}|^2}
\right)^{1/4}\exp\left(-\frac{(x-\tilde{x})^2}{2\tilde{\sigma}}-
\frac{\tilde{x}_i^2}{2\tilde{\sigma}_r}\right)\nonumber\\
\bar{x}&=&\tilde{x}_r+\frac{\tilde{\sigma}_i}{\tilde{\sigma}_r}\tilde{x}_i
\hspace{1cm}
\bar{p}=\hbar\frac{\tilde{x}_i}{\tilde{\sigma}_r}\nonumber\\
\overline{\Delta x^2}=
\frac{|\tilde{\sigma}|^2}{2\tilde{\sigma}_r} &&\overline{\Delta
p^2}=
\frac{\hbar^2}{2\tilde{\sigma}_r}\hspace{1cm}\overline{\Delta
x\Delta p+\Delta p\Delta
x}=\frac{\hbar\tilde{\sigma}_i}{\tilde{\sigma}_r}
\label{xx}\end{eqnarray} The values of these quantities will in
general depend on the value of $\theta$. In this subsection we will
supress this dependence but in the following we will denote the mean
position conditioned on a particular value of $\theta$ by
$\bar{x}_\theta$ and likewise for the other expectation values.  We
will now derive the dynamics of this 
state if a measurement takes place at time $\tau$. From time $0$
to $\tau^-$, just before the measurement, the equations of motion
are governed by the Schr\"odinger equation:
\begin{equation}\frac{d\tilde{\sigma}}{dt}=\frac{i\hbar}{m}
\left(1-\frac{m^2\omega^2}{\hbar^2}\tilde{\sigma}(t)^2\right)\hspace{1cm}
\frac{d\tilde{x}}{dt}=\frac{\tilde{\sigma}(t)}{i\hbar}\left(\theta+m
\omega^2\tilde{x}\right)\label{xx0}\end{equation}
The corresponding $\bar{x}$, $\bar{p}$ and second order moments
can easily be derived. The equation for $\tilde{\sigma}$ indicates
the spreading contracting of the wavepacket induced by the harmonic
oscillation. At
time $\tau$, the POVM $\{\hat{A}_{\xi}(\hat{x})\}$ is performed.
$\xi$ will be a Gaussian distributed random variable with
expectation value $\bar{x}(\tau^-)$ and variance
$D+\overline{\Delta x^2}(\tau^-)$. Straightforward calculations
show that the post-measurement wavefunction, conditioned on the
result $\xi$, is parameterized by:
\begin{equation}\frac{1}{\tilde{\sigma}(\tau)}=
\frac{1}{\tilde{\sigma}(\tau^-)}+\frac{1}{D}\hspace{1cm}
\tilde{x}_{\xi}(\tau)=\frac{\tilde{\sigma}(\tau^-)\xi+D\tilde{x}
(\tau^-)}{\tilde{\sigma}(\tau^-)+D}\label{sxi}\end{equation}
The equation for $\tilde{\sigma}$ now indicates the contracting effect of the
position measurement. The expectation values $\bar{x}$ and $\bar{p}$ become:
\begin{eqnarray}
\bar{x}(\tau)&=&\bar{x}(\tau^-)+\frac{|\tilde{\sigma}(\tau)|^2}
{\tilde{\sigma}_r(\tau)D}\left(\xi-\bar{x}(\tau^-)\right)\nonumber\\
\bar{p}(\tau)&=&\bar{p}(\tau^-)+\frac{\hbar\tilde{\sigma}_i(\tau)}
{D\tilde{\sigma}_r(\tau)}\left(\xi-\bar{x}(\tau^-)\right)\label{13}
\end{eqnarray}
Note that the back-action manifests itself by constantly
introducing white noise, i.e. $\xi-\bar{x}(\tau^-)$, into the
system.

It is trivial to write down the dynamical equations in the case
of a finite number ($N$) of  measurements: we just have to repeat
the previous two-stage procedure $N$ times. However we are
interested in taking the limit of infinitesimal time intervals
$dt$ between two measurements. This will only make sense if at
each infinitesimal time step the wavefunction is only subject to
an infinitesimal disturbance. Referring to equation (\ref{sxi}),
this implies that the measurement accuracy $D$ has to scale as
$1/dt$. Therefore we define the finite sensitivity $k$ by the
relation $D=1/(kdt)$, implying that only an infinitesimal amount
of information is obtained in each measurement. In this limit,
the random zero-mean variable $(\xi-\bar{x}(\tau^-))/D$ has a
standard deviation given by $\sqrt{kdt/2}$. This is very
convenient as a Gaussian random variable with zero mean and
variance $\sqrt{dt}$ is by definition a Wiener increment, and
therefore we can make use of the theory of Ito calculus. Defining
$d\xi(t)=\xi_tdt$ as being the measurement record, and using the
notation of Ito calculus, the complete equations of motion
conditioned on the measurement result for a Gaussian particle
subject to continuous observation of the position can be written
down:
\begin{eqnarray}
d\xi(t)&=&\bar{x}(t) dt+v_{\xi}(t)dW\label{17}\\
d\bar{x}(t) &=&\frac{\bar{p}(t)}{m}dt+v_x(t)dW\label{15}\\
d\bar{p}(t) &=&-m\omega^2\bar{x}(t) dt-\theta(t)dt+v_p(t)dW\\
\dot{\tilde{\sigma}}(t)&=&\frac{i\hbar}{m}\left(1-\frac{m^2\omega^2}{\hbar^2}\tilde{\sigma}(t)^2\right)-k(t)\cdot\tilde{\sigma}(t)^2\label{s1}\label{18}\\
v_x(t)=\sqrt{\frac{k(t)}{2}}\frac{|\tilde{\sigma}(t)|^2}{\tilde{\sigma}_r(t)}&&
v_p(t)=\sqrt{\frac{k(t)}{2}}\frac{\hbar\tilde{\sigma}_i(t)}{\tilde{\sigma}_r(t)}\hspace{1cm}
v_{\xi}(t)=\frac{1}{\sqrt{2k(t)}}\label{21}
\end{eqnarray}
If the sensitivity $k$ is kept constant during the whole
observation ($\forall t, k(t)=k(0)$), equation (\ref{s1}) can be
solved exactly. Given initial condition $\tilde{\sigma}_0$, the
solution is:
\begin{equation}
\tilde{\sigma}(t)=\tilde{\sigma}_{\infty}\left(\frac{\frac{\tilde{\sigma}_{\infty}+\tilde{\sigma}_0}{\tilde{\sigma}_{\infty}-\tilde{\sigma}_0}\exp(2i\Omega
t)-1}{\frac{\tilde{\sigma}_{\infty}+\tilde{\sigma}_0}{\tilde{\sigma}_{\infty}-\tilde{\sigma}_0}\exp(2i\Omega
t)+1}\right)\hspace{1cm} \Omega=\sqrt{\omega^2-\frac{i\hbar
k}{m}}\hspace{1cm}\tilde{\sigma}_{\infty}=\frac{\hbar/m}{\Omega}\label{steadys}
\end{equation}
This shows that the position variance of the wavefunction evolves at least
exponentially fast to a steady state. The damping is roughly proportional to
the square root of the sensitivity, while the steady state solution has a
variance inversely proportional to it. This result means that a continuously
observed particle is localized, although not confined, in space. It is
interesting to note that this localization increases with the mass of the
particle, such that it is very difficult to localize a light
particle. Indeed the steady state position variance can be understood
from the point of view of Standard Quantum Limits for position
measurement~\cite{Braginsky}. For example if $\omega^2 \gg \hbar k /m$ then
$\overline{\Delta x^2}_\infty \simeq \hbar /2m\omega$. Similarly, if we
take $t=1/\mathletters{Re}[\Omega]$ to be the time for an effectively
complete measurement, then for a free particle $\overline{\Delta
  x^2}_\infty = \hbar t /m$ and so the steady state position variance is the
same as the SQL for ideal position measurements separated by time
intervals of length $1/\mathletters{Re}[\Omega]$. \\

\subsection{Kalman filtering interpretation}
Let us now try to give a ``signal processing'' interpretation to equations
(\ref{17}-\ref{21}). The Wiener increment was defined as the difference
between the actual and the expected measurement result. As it is white noise,
it is clear that the expected measurement result was actually the best
possible guess for the result. This is reminiscent to the innovation process
in classical control theory: the optimal filtering equations of a classical
stochastic process can be obtained by imposing that the difference between
the actual and expected (i.e. filtered) measurement be white noise. Indeed,
in a previous paper \cite{Doherty1}, one of us noticed that the equations
(\ref{17}-\ref{21}) have exactly the structure of the Kalman filtering
equations associated with a classical stochastic linear system. This is in
complete accordance with the dynamical interpretation of quantum mechanics as
describing the evolution of our knowledge about the system.

The classical stochastic system that has exactly the same filtering equations
as our continuously observed quantum system is given by:
\begin{eqnarray}d\left(\begin{array}{c}x_{\theta}\\p_{\theta}\end{array}\right)&=&\left(\begin{array}{cc}0
&\frac{1}{m}\\-m\omega^2&0\end{array}\right)\left(\begin{array}{c}x_{\theta}\\p_{\theta}\end{array}\right)dt+\left(\begin{array}{c}0\\1\end{array}\right)\theta(t)dt+\left(\begin{array}{c}0\\
\hbar/2\end{array}\right)\sqrt{2k}dV_1 \nonumber\\
d\xi&=&\left(\begin{array}{cc}1&0\end{array}\right)\left(\begin{array}{c}x_{\theta}\\p_{\theta}\end{array}\right)dt+\frac{1}{\sqrt{2k}}dV_2\label{cse}\end{eqnarray}
$dV_1$ and $dV_2$ are two independent Wiener increments and
correspond to the process noise and measurement noise
respectively.  It is very enlightening to look at the
corresponding weights of these noise processes: the higher the
sensitivity, the more accurate the measurements, but the more
noise is introduced into the system. Moreover measuring the
position only introduces noise into the momentum. This clearly is
a succinct manifestation of the Heisenberg uncertainty relation.
Indeed, the product of the amplitude of the noise processes of
measurement and back-action is independent on the sensitivity $k$
and exactly given by $\hbar/2$.

The equations for the means $\bar{x}_{\theta}$ and $\bar{p}_{\theta}$ are now
given by the Kalman filter equations of this classical system, and the
equations for the variances $\overline{\Delta x^2_{\theta}} ,\overline{\Delta
p^2_{\theta}} ,\overline{\Delta x_{\theta}\Delta p_{\theta}+\Delta
p_{\theta}\Delta x_{\theta}}$ are given by the associated Riccati equations.
This is very convenient as this will allow us to use the convenient language
of classical control theory to solve the estimation problem.

\subsection{Continuous Parameter Estimation}
Let us now consider the basic question of this paper: how can we get the best
estimates of the unknown force $\{\theta(t)\}$ acting on the system, given
the measurement record $\{d\xi_t\}$? The natural way to attack this problem
is the use of Bayes rule. As we have a linear system with $\{d\xi_t\}$ a
linear function of $\{\theta(t)\}$, and the noise in the system is Gaussian,
this will lead to a Gaussian distribution in $\{\theta(t)\}$. Moreover, due
to the linearity, the second order moments of this distribution will be
independent of the actual measurement record. Therefore the accuracy of our
estimates will only be a function of the sensitivity chosen during the
observation process and of the prior knowledge we have about the signal
$\{\theta(t)\}$ (for example that it is constant) . This will allow us to
devise optimal measurement strategies.

The formalism that we have developed is particularly useful in the case that
we parameterize $\{\theta(t)\}$ as a linear combination of known
time-dependent functions $\{f_i(t)\}$, but with unknown weights
$\{\theta_i\}$:
\[\theta(t)=\sum_{i=1}^{n}\theta_i f_i(t)\]
The estimation, based on Bayes rule, will lead to a joint
Gaussian distribution in the parameters $\{\theta_i\}$. Indeed, we
have the relations:
\begin{eqnarray}
p(\{\theta_i\}|\{\xi(t+dt)\})&\sim&
p(d\xi(t)|\{\theta_i\},\{\xi(t)\})p(\{\theta_i\}|\{\xi(t)\})\nonumber\\
&\sim& p\left(d\xi(t)|
\bar{x}\left(t,\{\theta_i\},\{\xi(t)\}\right)\right)p(\{\theta_i\}|\{\xi(t)\})
\label{pxix}
\end{eqnarray}
In the last step we made use of the fact that the Kalman estimate
$\bar{x}_{\{\theta_i\}}(t)$ is a sufficient statistic for $d\xi(t)$. Moreover
all distributions are Gaussian, while $\bar{x}_{\{\theta_i\}}(t)$ is some
linear function of $\{\theta_i\}$ due to the linear character of the Kalman
filter:
\[\bar{x}_{\{\theta_i\}}(t)=\sum_i\theta_i \int_0^tdt'g(t,t')f_i(t')\] The function
$g(t,t')$ can easily be calculated using equations (\ref{17}-\ref{21}). To
obtain the variance of the optimal estimates of $\{\theta_i\}$, formula
(\ref{pxix}) has to be applied recursively. By explicitly writing out the
Gaussian distributions, and making use of the fact that the product of
Gaussians is still a Gaussian, it is then easy to show that the variances at
time $\tau$ are given by:
\begin{equation}
\frac{1}{\sigma_{\theta_i}}=\int_0^\tau
\frac{dt}{v_{\xi}^2(t)}\left(\int_0^t
dt'g(t,t')f_i(t')\right)^2\label{eoversa}
\end{equation}
A more intuitive way of obtaining the same optimal estimation, given a fixed
measurement strategy, of $\{\theta_i\}$ can be obtained by a little trick: we
can enlarge the state vector $(x_{\theta},p_{\theta})$ with the unknowns, and
construct the Kalman filter and Riccati equation of the new enlarged system.
$\bar{x}_{\theta}$ and $\bar{p}_{\theta}$, till now the expected values
conditioned on a fixed value of the force, then get the meaning of the mean
of these expected values over the probability distribution of the unknown
force. In other words, the new $\bar{x}$ and $\bar{p}$ become the ensemble
averages over the pure states labeled by a fixed force $\theta$. The new
enlarged system, in the case of one unknown parameter $\theta$, reads:
\begin{eqnarray}d\left(\begin{array}{c}x\\p\\ \theta\end{array}\right)&=&\underbrace{\left(\begin{array}{ccc}0&1/m&0\\-m\omega^2&0&f(t)\\0&0&0\end{array}\right)}_{A(t)}\left(\begin{array}{c}x\\p\\\theta\end{array}\right)dt+ \underbrace{\left(\begin{array}{c}0\\ \hbar/2\\0\end{array}\right)}_{B}\sqrt{2k(t)}dV1\\
d\xi&=&\underbrace{\left(\begin{array}{ccc}1&0&0\end{array}\right)}_{C}\left(\begin{array}{c}x\\p\\
\theta\end{array}\right)dt+\underbrace{\frac{1}{\sqrt{2k}}}_{D}dV_2\end{eqnarray}
The Kalman filter equations will give us the best possible
estimation of the vector $(x,p,\theta)$ at each time, while the
Riccati equation determines the evolution of the covariance
matrix $P$:
\begin{eqnarray}\frac{d}{dt}\left(\begin{array}{c}\bar{x}\\\bar{p}\\
\bar{\theta}\end{array}\right)&=&A(t)\left(\begin{array}{c}\bar{x}\\\bar{p}\\\bar{\theta}\end{array}\right)+2k(t)P(t)C^T\left(d\xi(t)-C\left(\begin{array}{c}\bar{x}\\\bar{p}\\
\bar{\theta}\end{array}\right)\right)\label{Kalric}\\
\dot{P}&=&A(t)P+PA^T(t)-2k(t)PC^TCP+2k(t)BB^T\label{Ric}\end{eqnarray} An
optimal measurement strategy, dependent on the sensitivity, will then be this
one that minimizes the $(3,3)$ element in $P$ at time $t_{final}$. An
analytic solution of this problem does not exist in general, as the Riccati
equations are quadratic. However, in the case of constant $f(t)=f(0)$ and
constant sensitivity $k(t)=k(0)$ analytical results will be derived.

Before proceeding however, it is interesting to do a dimensional analysis to
see how the variances will scale. We begin by scaling $\tilde{t}=t/\tau$ with
$\tau$ the duration of the complete measurement. Introducing the matrix
\begin{equation}T=\left(\begin{array}{cccc}\sqrt{\frac{\hbar
\tau}{2m}}&0&0\\0&\sqrt{\frac{\hbar
m}{2\tau}}&0\\0&0&\sqrt{\frac{\hbar
m}{2\tau^3}}\end{array}\right),\end{equation} it can easily be
checked that $\tilde{P}=T^{-1}PT^{-1}$ is dimensionless. If we
then scale the sensitivity as $k(t)=\tilde{k}(\tilde{t})\hbar
\tau^2/(2m)$, the force $\theta=\tilde{\theta}\sqrt{\hbar
m/2\tau^3}$ and do the appropriate transformations
$B\rightarrow\tilde{B}$ and $C\rightarrow\tilde{C}$, we get the
equivalent state space model:
\begin{equation}\tilde{A}=\left(\begin{array}{ccc}0&1&0\\-\omega^2\tau^2&0&f(t)\\0&0&0\end{array}\right)\hspace{1cm}\tilde{B}=\left(\begin{array}{c}0\\1\\0\end{array}\right)\hspace{1cm}\tilde{C}=\left(\begin{array}{ccc}1&0&0\end{array}\right)\label{dimens}\end{equation}
The new filter equations are still given by (\ref{Kalric},\ref{Ric}) with the
substitution
$(A,B,C,k(t))\rightarrow(\tilde{A},\tilde{B},\tilde{C},\tilde{k}(\tilde{t}))$.
This observation has an immediate consequence if we are measuring the force
acting on a free particle ($\omega=0$): the standard deviation on our
estimate will always scale like $\sqrt{\hbar m/2\tau^3}$, and the chosen
sensitivity will only affect the accuracy by a multiplicative pre-factor.

\section{General formalism for quantum parameter estimation}

In this section we develop a description of the problem of estimating unknown
parameters $\theta$ of the dynamics of a quantum system from the results of
generalized measurements. This general problem can be addressed in
essentially the same way as the specific problem of force estimation
for an oscillator that was discussed in the previous section. An
approach to this problem has been 
proposed by one of us~\cite{mabuchi1996b} and we will formulate the theory in
the language of operations and effects and consider in particular the case of
measurement currents that are continuous in time, as in the case of homodyne
detection~\cite{wiseman1993a} or continuous position measurement. The
fundamental basis of this approach is to propagate an \emph{a
  posteriori} probability distribution 
$p\left( \theta  | {\mathbf I}_{[0,t)}\right) $ for the parameter
$\theta$ conditioned 
on the history of measurement results ${\mathbf I}
_{[0,t)}$ up to time $t$ by 
employing Bayes' rule and using the theory of operations and effects to
calculate the relative likelihood of the known measurement record as a
function of $\theta$. Readers who are less interested in mathematical details
and more interested in the application of our formalism to the force
estimation problem may skip this section.

\subsection{General Theory}
We will treat the quantum parameter estimation as an essentially classical
parameter estimation problem coupled to the quantum measurement updating
rules. For each value $\theta '$ of $\theta $ there will be a
conditioned state $\rho 
_{\theta '}$ describing the state of the quantum system conditioned on the
measurement history and a particular value of the unknown parameter $\theta
$. This density matrix would be our best description of the state if we knew
the measurement record and also that $\theta $ took this particular value.
However the value of $\theta$ is not assumed to be known exactly and is
described by a probability distribution $p(\theta)$. Hence the density matrix
describing the state from the point of view of the experimenter is
\begin{equation}
\rho=\int d\theta p(\theta) \rho_\theta. \label{defn}
\end{equation}

The most general quantum evolution and measurement can be described by the
theory of operations and effects. The following discussion will adapt the
treatment of Wiseman and Diosi to our problem~\cite{wiseman2000a}. In this
work we assume that either the dynamics or the measurement are unknown and
belong to a family parameterized by $\theta$. Thus we consider quantum
measurements characterized by a set of operators $\Omega_{\theta, r}$ where
$\theta$ labels the value of the unknown parameter and $r$ labels the
measurement result. Thus there is a separate measurement for each value of
$\theta$ and the operators $\Omega_{\theta, r}$ are constrained by
completeness
\begin{equation}
\int d\mu_{\theta,0}(r)\Omega_{\theta, r}^\dagger \Omega_{\theta,
r}=1. \label{eq:complete}
\end{equation}
Here $d\mu_{\theta,0}(r)$ is a normalized measure on the space of
measurement results $r$. As in the standard theory, the
probability of the measurement result $r$ conditioned on $\theta$
is
\begin{equation}
d\mu_{\theta}(r)=
d\mu_{\theta,0}(r)\text{Tr}\left[\Omega_{\theta, r}^\dagger
  \Omega_{\theta, r}\rho_{\theta}\right].
\end{equation}
The state of the quantum system after the measurement conditioned on the pair
$(\theta,r)$ is
\begin{equation}
\label{eq:condstate}
\rho_{\theta,r}'=\frac{d\mu_{\theta,0}(r)\Omega_{\theta, r}
  \rho_{\theta} \Omega_{\theta, r}^\dagger}{d\mu_{\theta}(r)} =
  \frac{\Omega_{\theta, r}
  \rho_{\theta} \Omega_{\theta,
  r}^\dagger}{\text{Tr} \left[\Omega_{\theta, r}^\dagger
  \Omega_{\theta, r}\rho_{\theta} \right] }.
\end{equation}
If the result of the measurement is unknown or disregarded then
the state of the system is an average over the conditioned states
weighted by their probabilities
\begin{equation}
\rho_\theta'=\int d \mu_\theta(r )\rho_{\theta,r}' = \int
  d\mu_{\theta,0}(r) \Omega_{\theta, r}
  \rho_{\theta} \Omega_{\theta,r}^\dagger.
\end{equation}
This is the state of the system conditioned on a particular value
of $\theta$ but not on any measurement result.

The unconditioned probability of the measurement results is found
by averaging over the probability distribution for $\theta$ and
is given by the measure
\begin{equation}
d\mu(r)= \int_\theta d\theta p(\theta) d\mu_{\theta}(r)
=\int_\theta d\mu(\theta) d\mu_{\theta}(r).
\end{equation}
After the measurement we will require that the state conditioned
on the measurement result $r$ but not on  the value of $\theta$ may
still be written in the form of 
Eqn.\ (\ref{defn}) as an average over the states conditioned on
particular values of $\theta$, thus
\begin{equation}
\rho_r '=\int d \mu_r(\theta )\rho_{\theta,r}'
\end{equation}
for some measure $d \mu_r(\theta )$ on the space of possible
$\theta$. This new measure describes the probability of $\theta$
conditioned on the measured value of $r$. This conditioned
probability distribution for $\theta$ is precisely what we wish
to calculate. For consistency it must be the case that if the
measurement result is unknown or disregarded the appropriate
state is again an average over the conditioned states
\begin{equation}
\rho '=\int d \mu(r )\rho_{r}' = \int
  d\mu(r) d\mu_{r}(\theta) \rho_{\theta,r}'.
\end{equation}

In order to calculate $d \mu_r(\theta )$ we need to develop a
Bayes rule that relates all the probability measures we have
introduced. In order to do this we note that $\rho'$ must also be
able to be expressed as an average over the probability for
$\theta$ of the states $\rho_\theta '$, thus
\begin{equation}
\rho '=\int d \mu(\theta)\rho_{\theta}' = \int
  d\mu(\theta) d\mu_{\theta}(r) \rho_{\theta,r}'.
\end{equation}
This leads us to the Bayes rule
\begin{equation}
\label{eq:bayes} d\mu(r)d\mu_{r}(\theta)= d\mu(\theta)
d\mu_{\theta}(r)
\end{equation}
which allows us to calculate $d\mu_{r}(\theta)$ in terms of
$d\mu(\theta)$, the measure that characterizes our prior knowledge
about $\theta$, and the measures $d\mu_{\theta}(r)$, which are
part of our specification of the parameterized family of
measurements.

In principle this allows us to optimally update the probability
distribution for the unknown parameter in any quantum
measurement. We are most interested here in the case of
measurements that are continuous in time. In this situation we
wish to derive a stochastic differential equation that updates
the distribution for $\theta$ conditioned on measurement current.
Since the case of photon detection style measurements is
considered in~\cite{mabuchi1996b} we will consider measurements
like homodyne detection where the measurement results are
continuous but not differentiable functions of time,
in~\cite{wiseman2000a} these are termed diffusive measurements.
This will require that we develop stochastic differential
equations to describe the measurement process.

For simplicity we will consider the case where there is only a
single measurement being made and we will describe the
measurement result $r$ in an infinitesimal time interval
$[t,t+dt)$ by the complex number $I(t)$. We define the
measurement operators
\begin{equation}
  \label{eq:operation}
  \Omega_{\theta,I} = 1-iH(\theta)dt-\frac{1}{2}\hat{c}^\dagger
  \hat{c} +I^*\hat{c}dt.
\end{equation}
These measurement operators may be derived, for example, as the
continuous limit of a model of repeated
measurements~\cite{caves1987a} or from models of quantum optical
measurements such as heterodyne or homodyne detection \cite{wiseman1993a}. For
simplicity we consider the case where there is only one
measurement current, the general case may easily be treated
following the formalism of~\cite{wiseman2000a}. We also assume
that the specific measurement that is being made is known (that
is that the operator coupling the system to the bath and the
measurement made on the bath are known) and so $\theta$ only
parameterizes the Hamiltonian evolution of the system. This is
the most interesting case and simplifies the treatment. The
extension to the case where the measurement is known but the free
system evolution is not unitary but is rather described by a
Markovian master equation is also straightforward. Now the
measurement operator is constrained by the completeness relation
Eq.\ (\ref{eq:complete}) and this requires that
\begin{eqnarray}
  \label{eq:measure}
  \int d\mu_{\theta,0}(I) (Idt) & = & 0  \\
  \int d\mu_{\theta,0}(I) (I^*dt)(Idt) & = & dt.
\end{eqnarray}
These moments mean that we may identify $Idt$ as a complex Wiener
increment under the measure $d\mu_{\theta,0}(I)$. However in
order to specify this measure completely we must also specify the
remaining second order moment of the Wiener increment (clearly
this must also be of order $dt$). We will say that
\begin{equation}
  \label{eq:measure2}
  \int d\mu_{\theta,0}(I) (Idt)(Idt)  =  udt,
\end{equation}
where we need $|u|\leq1$. In line with our assumption that the
measurement interaction and the measurement on the bath is known
we will require that $u$ is independent of $\theta$. The case
$u=0$ corresponds in the quantum optical setting to heterodyne
detection, while $|u|=1$ corresponds to homodyne detection with
some local oscillator phase. Note that these moments are
independent of $\theta$ and so we can drop the subscript $\theta$
for this measure on $I$ from here on. Since the moments of $Idt$
under $d\mu_0(I)$ indicate that we consider $Idt$ to be a complex
Wiener increment, we adopt the Ito rules
\begin{equation}
  \label{eq:wiener}
  (Idt)^2=udt,\quad(I^*dt)(Idt)=dt.
\end{equation}

Now we would like to know the observed statistics of $I$ under the
physical measure $d\mu(I)$. There are two kinds of conditioned
expectation values for operators $\hat{a}$ in this problem.
Expectation values conditioned on a particular value of the
unknown parameter will be denoted $\bar{a}_{\theta
}=$Tr$[\hat{a}\rho _{\theta }]$. On the other hand expectation
values
conditioned only on the history of measurement results will be denoted $%
\bar{a} =$Tr$[\hat{a}\rho ]$. Now we know from the preceding
discussion that
\begin{eqnarray}
  d\mu(I)&=&\int_\theta d\mu(\theta) d\mu_0(I) \text{Tr}
  \left[\Omega_{\theta, I}^\dagger \Omega_{\theta, I}\rho_{\theta}
  \right]\\
  &=&  d\mu_0(I) \int_\theta d\mu(\theta) \text{Tr}\left[(1+I^*\hat{c}
  dt
  +I \hat{c}^\dagger dt) \rho_{\theta} \right] \\
  &= & d\mu_0(I) (1+I^*dt  \bar{c} +Idt
  \bar{c}^\dagger ). \label{eq:physmeas}
\end{eqnarray}
Hence the expected value of $I$ is
\begin{equation}
  \label{eq:mean}
  \langle I \rangle = \int d\mu(I) I = u\bar{c}^\dagger +\bar{c}.
\end{equation}
From Eq.\ (\ref{eq:physmeas}) we can see that the second order
moments of $Idt$ are independent of the state and of $\theta$ and
are equal to the second order moments under $d\mu_0(I)$. Thus the
transformation from the measure $d\mu_0(I)$ to $d\mu(I)$ is a
transformation of drift similar to a Girsanov
transformation~\cite{Oksendal} and
we can identify $Idt$ with
\begin{equation}
  \label{eq:meascurrent}
  Idt=u\bar{c}^\dagger +\bar{c} dt+dW
\end{equation}
where $dW$ is a complex Wiener increment under the measure $d\mu
(I)$ obeying $dW^2=udt, dW^*dW=dt$.

On the other hand the probability measure for the measurement
trajectories conditioned on a given value of $\theta$ is
\begin{eqnarray}
  d\mu_\theta(I)&=& d\mu_0(I) \text{Tr}
  \left[\Omega_{\theta, I}^\dagger \Omega_{\theta, I}\rho_{\theta}
  \right]\\
  &= & d\mu_0(I) (1+I^*dt \bar{c}_\theta +Idt
  \bar{c}^\dagger_\theta ). \label{eq:condmeas}
\end{eqnarray}
Using Eq.\ (\ref{eq:bayes}) it is now straightforward (keeping
terms up to second order in $Idt$) to update the probability for
$\theta$ conditioned on $I$
\begin{equation}
  \label{eq:update}
  d\mu_{{\mathbf I}_{[0,t+dt)}} = \left[1+\left( \bar{c}_{\theta}-\bar{c}
       \right) \left( 
I^*dt-u^*\bar{c}dt -\bar{c}^{\dagger } dt\right)+\left(
\bar{c}^{\dagger}_{\theta}-\bar{c}^\dagger  \right) \left( 
Idt-\bar{c}dt -u\bar{c}^{\dagger } dt\right)
\right]d\mu_{\mathbf{I}_{\mathnormal{[0,t)}}}(\theta).
\end{equation}
This allows us to write down a stochastic Fokker-Planck equation
for the probability distribution of $\theta$
\begin{equation}
  dp(\theta)|{\mathbf I}_{[0,t+dt)})=\left[ \left( \bar{c}_{\theta}-\bar{c}
       \right) \left( 
I^*dt-u^*\bar{c}dt -\bar{c}^{\dagger } dt\right) + \mathletters{H.c.}
\right] p(\theta|{\mathbf I}_{[0,t)})
\label{QPE} 
\end{equation}
Note that under $d\mu(I)$ the innovation
$Idt-\bar{c}dt-u\bar{c}^{\dagger
  }dt$ is a Wiener increment and thus has mean zero and is not
correlated with either the quantum state or $p(\theta)$. This
  equation is very similar in form to the Kushner-Stratonovich
  equation that arises in classical state estimation
  problems~\cite{maybeck1982a}. In order to
be able to propagate this equation for the probability
distribution of $\theta$ we must also be able to update the
conditioned state $\rho_{\theta}$ and hence the expectation
values $\bar{c}_\theta$. From
  Eq.\ (\ref{eq:condstate}) we can show that
$\rho_{\theta}$ obeys the stochastic master equation (SME)
\begin{equation}
d\rho _{\theta }= -i[H\left( \theta \right), \rho _{\theta
  }]dt+{\mathcal D}[\hat{c}]\rho _{\theta }dt+{\mathcal H}\left[ \hat{c}
  \left( I^*dt-
  \bar{c}^\dagger dt-u^*\bar{c}_\theta dt\right) \right] \rho
_{\theta } . \label{SME}
\end{equation}

Equation (\ref{QPE}) and the family of stochastic master
equations (\ref{SME}) describe the quantum parameter estimation
problem for measurements with continuous measurement currents
such as optical homodyne detection. As we indicated at the start
of this section, and as in the algorithm discussed
in~\cite{mabuchi1996b}, a family of quantum states conditioned on
the measurement record and on different values of $\theta$ is
propagated using appropriate SME's while the conditioned
probability distribution for $\theta$ is propagated using a
stochastic Fokker-Planck equation of the kind that arises in
classical estimation problems. As we shall see below it is
possible to solve these equations for certain linear models such
as force estimation due to position measurement on a free
particle or oscillator. In general it will be necessary to
integrate these equations numerically after first discretizing
$\theta$. In principle this is straightforward although the
discretization must be sufficiently fine that a good
approximation for the mean $\bar{c}+u\bar{c}^{\dagger } $ is
maintained at all times and this will usually involve a prohibitive
computational cost. One way of avoiding this is to consider a
linear variant of this update equation which is in fact more
closely allied to the algorithm in~\cite{mabuchi1996b}. This
variant is an analogue both of the linear version of the
stochastic master equation~\cite{goetsch1994a} and of the Zakai
equation which is the linear counterpart to the
Kushner-Stratonovich equation~\cite{maybeck1982a} in classical
state estimation. This linear variant does not preserve the
normalization of $p\left( \theta |I\right) $ but does not depend
on $u\bar{c}+\bar{c}^{\dagger }$ and yet still propagates the
relative probabilities of different values of $\theta$.

The basic observation is that in the Bayes' rule Eq.\
(\ref{eq:bayes}) the measure $d\mu(r)$ is independent of $\theta$
and only ensures the normalization of $d\mu_r(\theta)$. If we are
only interested in the relative likelihood of different values of
$\theta$ we may consider unnormalized measures
$d\bar{\mu}_r(\theta)$ on the space of possible $\theta$ and
replace $d\mu(r)$ by any measure on $r$ independent of $\theta$.
In particular for our example of continuous measurements we may
choose
\begin{equation}
  \label{eq:unnorm}
  d\bar{\mu}_{{\mathbf I}_{[0,t+dt)}}(\theta) d\mu_0 (I) = d \mu_\theta
  (I) d\bar{\mu}{\mathbf I}_{[0,t)} (\theta). 
\end{equation}
Substituting from Eq.\ (\ref{eq:condmeas}) we get
\begin{equation}
d\tilde{p}\left( \theta |{\mathbf I}_{[0,t+dt)}\right) =\left( \bar{c}_\theta
I^*dt
  + \bar{c}^{\dagger
} _{\theta }Idt\right) \tilde{p}\left( \theta |{\mathbf I}_{[0,t+dt)}\right) .
\label{QPE2}
\end{equation}
Under this linear propagation equation the dynamics of the
unnormalized distribution $\tilde{p}\left( \theta
  |{\mathbf I}_{[0,t)} \right) $ may be calculated for each value of
$\theta $ independently. This will make it possible to calculate
relative probabilities of a discrete set of possible values of
$\theta $ given a particular sequence of measurement results with
no constraints on the discretization of $\theta $.

This formalism for the estimation of a classical parameter in
quantum dynamics may readily be generalized to the case where there is
more than one unknown parameter or where the parameter undergoes some
known time dependence as in the previous section. Another
interesting situation that may be treated straightforwardly in this
formalism is correlating the measurement results from
two quantum measurements both of which depend on $\theta$. Here
we have assumed that apart from the measurement the dynamics of
the quantum system is unitary. If this is not true (as is the
case for less than perfectly efficient detection for example)
then it is straightforward to show that the first term of Eq.\
(\ref{SME}) is simply replaced by a Liouvillian term describing
the noisy dynamics of the system, thus
\begin{equation}
d\rho _{\theta }= {\mathcal L}(\theta) \rho _{\theta
  }dt+{\mathcal D}[\hat{c}]\rho _{\theta }dt+{\mathcal H}\left[ \hat{c}
  \left( I^*dt-
  \bar{c}^\dagger_\theta dt -u^*\bar{c} _\theta dt\right) \right] \rho
_{\theta} . \label{SME2}
\end{equation}
In the next section we will return the problem of force estimation
through continuous position measurement of an oscillator. We will be
most interested in finding the optimum (possibly time-dependent)
sensitivity of the measurement.

\subsection{Force Estimation through Continuous Position Measurement}

The general formalism of this section may be reduced to the
parameter estimation problem we considered at the start of the
paper in the important case of force estimation though continuous
position measurement of an oscillator
($\hat{c}=\sqrt{2k}\hat{x},u=1,H(\theta)=\hat{p}^2/2m+m\omega^2\hat{x}^2/2
+\theta \hat{x}$). In this case it is possible to solve the system
of equations (\ref{SME}) and (\ref{QPE}) explicitly. We have the
system of equations
\begin{eqnarray}
  \label{eq:forceest}
  d\rho _{\theta }&=& -i[\hat{p}^2/2m+m\omega^2\hat{x}^2/2
+\theta \hat{x},
  \rho _{\theta
  }]dt+2k{\mathcal D}[\hat{x}]\rho _{\theta
  }dt+\sqrt{2k}{\mathcal H}\left[ \hat{x}
   \right] \rho
_{\theta }\left(
Idt-2\sqrt{2k}\bar{x}_\theta dt \right)  \\
dp(\theta|{\mathbf I}_{[0,t+dt)})&=&2\sqrt{2k}\left( \bar{x}_{\theta}-\bar{x}
  \right) \left(
Idt-2\sqrt{2k}\bar{x}dt \right) p(\theta|{\mathbf I}_{[0,t+dt)}).
\end{eqnarray}
This linear system preserves Gaussian quantum states of the
oscillator and Gaussian probability distributions for $\theta$.
As a result we only need to find stochastic equations for the
first and second order moments of the $\rho_\theta$ and
$P(\theta|I)$. The procedure is to apply standard master equation
techniques~\cite{walls1994a} combined with the Ito rules for
stochastic differential equations to find equations for the
moments of $\hat{x}$ and $\hat{p}$, conditioned on a particular
value of $\theta$, from Eq.~(\ref{eq:forceest}) as was done
in~\cite{Doherty1}. The unconditioned moments result from
averaging over $p(\theta|I)$
\begin{eqnarray}
  \label{eq:moments}
  \overline{\Delta x^{2}} &=& \int d\theta p(\theta)
  \text{Tr}\left[ \left(\hat{x}-\bar{x}_\theta\right)^2 \rho_\theta \right] \\
  \overline{\Delta x\Delta p} &=& \int d\theta p(\theta)
  \left( \text{Tr}\left[(\hat{x}\hat{p}+\hat{p}\hat{x})\rho_\theta
  \right]/2 - \bar{
  x}_\theta\bar{p}_\theta \right)  \\
  \overline{\Delta p^{2}} & =&\int d\theta p(\theta)
  \text{Tr}\left[ \left(\hat{p}-\bar{p}_\theta\right)^2 \rho_\theta \right] \\
  \overline{\Delta x \Delta \theta} &=& \left(\int d\theta p(\theta)
  \theta \bar{x}_\theta\right) - \bar{\theta}  \bar{ x} \\
  \overline{ \Delta p \Delta \theta } &=& \left( \int d\theta p(\theta)
  \theta \bar{p}_\theta \right)- \bar{\theta} \bar{p}\\
  \overline{\Delta \theta ^{2}} &=& \int d\theta
  p(\theta)\left(\theta -\bar{\theta}\right)^2.
\end{eqnarray}
These moments form the covariance matrix $P$ and it is a
straightforward though tedious exercise to show that it obeys the
matrix Riccati equation (\ref{Ric}) we derived in the first section.
Similarly the first order moments obey the equations (\ref{Kalric}) of the
Kalman estimator.

\section{Standard Quantum Limits}
The preceding sections dealt with the problem of optimal
estimation of parameters of the Hamiltonian given a system that
is continuously observed. In this section we will derive the
explicit equations of the variances on these estimates.

\subsection{Detection of stationary signals}
Let us first introduce the idea of the standard quantum limit in
the context of von Neumann measurements. The idea is that a
particle is prepared in some optimal way at time $0$, such that
at time $\tau$ a projective measurement is performed to determine
the displacement associated with the force. The optimal
preparation is crucial as it has to balance the position and the
momentum uncertainty. The optimal preparation leads to the
expression of the Standard Quantum Limit. Consider a free
particle with a Gaussian wavefunction $\langle x|\psi\rangle$ and
initial parameters $\tilde{x}(0),\tilde{\sigma}(0)$ (see equation
(\ref{xx})) and subject to an unknown force $\theta$. The
integrated equations of motion (\ref{xx0}) are given by:
\[\tilde{x}(t)=\tilde{x}_0+\theta\left(t\tilde{\sigma}
(0)/i\hbar+t^2/2m\right)\hspace{1cm}\tilde{\sigma}(t)=\tilde{\sigma}(0)+i\frac{\hbar}{m}t\]
Suppose that at time $\tau$ we perform a von Neumann measurement
of the position. The probability distribution associated with this
measurement is given by:
\begin{equation}
p(x|\theta)\sim\exp\left(-\frac{\left(x-\frac{\theta
t^2}{2m}\right)^2}{|\tilde{\sigma}|^2/\tilde{\sigma_r}}\right)
\end{equation}
Using Bayes rule assuming a flat prior distribution for $\theta$ the variance
on the estimate of $\theta$ given the measurement result $x$ can
easily be derived:
\begin{equation}\sigma_\theta=\frac{2m^2|\tilde{\sigma}(t)|^2}
{\tilde{\sigma}_r(t)t^4}=\frac{2m^2\left(\tilde{\sigma}_r^2(0)+(\tilde{\sigma}_i(0)+\frac{\hbar
t}{m})^2\right)}{\tilde{\sigma}_r(0)t^4}\end{equation} This function is
heavily dependent on the initial conditions of the wavefunction of the
particle. The standard quantum limit can now be derived by choosing the
initial conditions such that $\sigma_\theta$ is minimized. This variance can
in principle go to zero if we allow $\langle \Delta
x\Delta p\rangle $ to be negative, but we will not consider such
``contractive'' states \cite{Yuen,Ozawa} here.  We therefore impose the
condition $\tilde{\sigma}_i(0)\geq0$ in order to focus our attention on the
specific issue of sensitivity optimization.  The optimal $\tilde{\sigma}(0)$
is then given by $\tilde{\sigma}(0)=\hbar t/m$, and this leads to the
expression of the Standard Quantum Limit:
\begin{equation}\sigma_\theta=\frac{4\hbar m}{t^3}\label{SQLclas}\end{equation} It is
clear that the square of the amplitude of a detectable force has to be bigger
than the variance on its estimation to be detectable. Therefore
the previous formula is the expression of the minimal force that
can be detected by a free particle of mass $m$ over a time $t$.
Note that the derived formula exceeds the normal equation of the
SQL by a factor 8 as the standard equation is not derived in the
context of parameter estimation.

We will now apply an analogous reasoning to a quantum particle
subject to continuous measurement. The explicit expression of the
variance on the estimated force was given by equation
(\ref{eoversa}). As noted at the end of the first section, the
resulting variance will be given by the standard quantum limit
multiplied by a certain factor. From here on we will therefore
work in the dimensionless picture as defined in (\ref{dimens}).
In general it is very hard to find the explicit expression for
the autocorrelation function $g(t,t')$ in equation
(\ref{eoversa}). Things get much more feasible if we do not vary
the sensitivity during the measurement as the system then becomes
stationary. It follows that we can assume that the values of the
variances reached their steady state values given by equation
(\ref{steadys}). After some straightforward linear algebra, the
explicit expression for $g(t,t')$ in the case of steady state is
given by:
\begin{eqnarray}
g(t,t')&=&\frac{1}{b}\exp(-a(t-t'))\sin(b(t-t'))\\
a&=&\omega\tau\sqrt{\frac{1}{2}\left(-1+\sqrt{1+\frac{(2k)^2}{(\omega\tau)^4}}\right)}\\
b&=&\omega\tau\sqrt{\frac{1}{2}\left(1+\sqrt{1+\frac{(2k)^2}{(\omega\tau)^4}}\right)}\label{gttt}
\end{eqnarray}
Due to the stationarity of the variances, the autocorrelation
function $g(t,t')$ is indeed only dependent on $(t-t')$, and from
here on we will therefore use the notation $g(t,t')=g(t-t')$. The
full expression of the variance on our estimate now becomes:
\begin{equation}\frac{1}{\sigma_\theta}=2k\int_0^1dt\left(\int_0^t
dt'g(t-t')f(t')\right)^2\label{blasa}\end{equation} The force that
acted on the system was assumed to be of the form
$\theta(t)=\theta f(t)$ with $f(t)$ a known function. Note that
this expression is dimensionless and has to be multiplied by
$\frac{2\tau^3}{\hbar m}$. We next introduce $F(\omega)$ and
$G(\omega)$ the Fourier transforms of the functions $f(t)\cdot
u_{[0,1]}(t)$ and $g(t)\cdot u_{[0,1]}(t)$, where $u_{[0,1]}(t)$
is the window function over the interval $[0,1]$. The damping
effect due to the back-action noise is responsible for broadening
the spectrum of the harmonic oscillator with a width of
approximately $k/(\omega\tau)$. Basic properties of Fourier
transformations lead to the expression:
\begin{equation}\frac{1}{\sigma_\theta}=\frac{2k}{(2\pi)^2}\int_{-\infty}^\infty\int_{-\infty}^\infty
d\omega_1 d\omega_2
\exp\left(i\frac{\omega_1-\omega_2}{2}\right)\frac{\sin\left(\frac{\omega_1-\omega_2}{2}\right)}{\frac{\omega_1-\omega_2}{2}}
G(\omega_1)G^*(\omega_2)F(\omega_1)F^*(\omega_2)\label{centrsa}\end{equation}
This formula clearly shows that only the frequencies of the signal
$F(\beta)$ near to the natural frequencies of the oscillator
$G(\beta)$ will be detectable.

Now we shall explicitly calculate the value of $\sigma_\theta$ in
some different cases. Let us first of all assume that the
spectrum $F(\beta)$ is almost constant for all values where
$G(\beta)$ is substantially different from $0$, i.e. around
$\beta\simeq (\omega\tau)$. This is realistic in some scenarios of
interest for the detection of gravitational waves
\cite{Braginsky}. Let us furthermore assume that $\omega\tau\gg
1$, which means that the period of the oscillator is much smaller
then the observation time. Next we observe that we are allowed to
approximate the ${\rm sinc}((\omega_1-\omega_2)/2)$ function by a
delta-Dirac function if the width of the spectrum $G(\beta)$,
determined by the number $k/(\omega\tau)$, is much bigger then
$1$. This leads to the expression:
\begin{eqnarray}\frac{1}{\sigma_\theta}&\simeq&\frac{k|F(\omega\tau)|^2}{2\pi}\int_{-\infty}^\infty
d\omega |G(\omega)|^2\\
&=&
\frac{k|F(\omega\tau)|^2}{2\pi}\int_{-\infty}^\infty d\omega\frac{1}{(a^2+b^2-\omega^2)^2+4a^2b^2}\\
&=&\frac{|F(b)|^2}{4\omega\tau}\chi\left(\frac{2k/(\omega\tau)^2}{\sqrt{1+(2k/(\omega\tau)^2)^2}}\right)\\
\chi(x)&=&(1-x^2)^{1/4}\sqrt{\frac{1+\sqrt{1+x^2}}{2(1+x^2)}}\end{eqnarray}
The function introduced in the last line is only dependent on
$2k/(\omega\tau)^2$, which can be tuned freely by changing the
value of our sensitivity. The function $\chi(x)$ reaches its
maximum value $1$ for small values of $x$, meaning that optimal
detection requires $k\ll (\omega\tau)^2$. The derivation however
required that $1\ll k/(\omega\tau)$. Therefore, the optimal choice
of the sensitivity will be given by a value $(\omega\tau)\ll
k\ll(\omega\tau)^2$, leading to the variance on the estimate:
\begin{equation}\sigma_\theta^2\simeq \frac{4\omega\tau}{|F(b)|^2}\frac{\hbar
m}{2\tau^3}=\frac{1}{|F(b)|^2}\frac{2\hbar
m\omega}{\tau^2}\end{equation} This corresponds exactly to the
expression of the standard quantum limit for an oscillator
\cite{Braginsky}. A similar expression can be obtained by
explicitly integrating (\ref{blasa}) with $f(t)=\delta(t)$. The
conditions under which this SQL can be reached are: 1. The total
duration of the measurement is much bigger then the period of the
oscillator; 2. The spectrum of the signal to be detected is flat
around the natural frequencies of the observed oscillator.

We will now investigate what happens if this second condition is
not fulfilled. In the extreme case, the force to be detected is
constant, corresponding to a delta-Dirac function in the frequency
domain. Again under the condition that $1\ll \omega\tau\ll
k/(\omega\tau)$, a good approximation of equation (\ref{centrsa})
becomes: \begin{equation}\frac{1}{\sigma_\theta}\simeq
k|G(0)|^2=\frac{1}{(\omega\tau)^2}\frac{2k/(\omega\tau)^2}{1+\left(2k/(\omega\tau)^2\right)^2}\end{equation}
The optimal sensitivity is now given by $2k=(\omega\tau)^2$,
indicating that one has to choose a much higher sensitivity to
detect constant forces then resonant oscillating forces. The
expression for the SQL for detecting constant forces with a
harmonic oscillator therefore becomes:
\begin{equation}\sigma_\theta\simeq 2(\omega\tau)^2\frac{\hbar
m}{2\tau^3}=\frac{m\hbar\omega^2}{\tau}\end{equation}

It is now natural to look what happens in the limit of
$\omega\rightarrow 0$, it is if the observed particle is free and
only subject to a constant force. In that case the explicit
integration of (\ref{gttt}) becomes possible, as $a$ and $b$ both
become equal to the sensitivity $\sqrt{k}$. Straightforward but
long integrations lead to:
\begin{equation}\sigma_\theta=\frac{8k^{3/2}}{4\sqrt{k}-5+8\exp(-\sqrt{k})\cos(\sqrt{k})-\exp(-2\sqrt{k})
\left(2+\cos(2\sqrt{k})+\sin(2\sqrt{k})\right)}\end{equation}
Minimization over the sensitivity leads to an expression for the
SQL for the detection of a constant force with a free particle
subject to continuous observation:
\begin{equation}\sigma_\theta\simeq 3\frac{4\hbar m}{\tau^3}.\end{equation} Note
that this expression differs from the corresponding one derived in
\cite{mabuchi1}, where calculations were done without properly accounting for
the damping effect of measurement back-action.  Comparing this result with
(\ref{SQLclas}), the variance of our estimate obtained by continuous
measurement is 3 times bigger then if we were doing projective measurements.
This is caused by two factors: at the end of the continuous measurement,
there is still a lot of information encoded about the force in the
wavefunction as the variance on the position at time $\tau$ is not at all
equal to $\infty$. Secondly, the previous result was obtained by assuming
that the variances of our Gaussian wavefunction were in steady state, and
this is not necessarily the optimal initial condition. Indeed, it turns out
that the optimal initial state (not considering contractive states) of the
continuously observed particle is a Gaussian state with well defined momentum
($\langle\Delta p^2\rangle\ll 1$) and therefore undefined position
$\langle\Delta x^2\rangle\gg 1$. This makes sense as the force to be detected
can only be seen because it manifests itself through the momentum. The fact
that the position uncertainty is very large is not so bad as the position is
continuously observed such that it becomes well-defined very quickly. The
expression for the variance on the force estimate using this optimally
prepared initial state can now be calculated exactly by explicitly solving
the Riccati equations (\ref{Ric}):
\begin{equation}\sigma_\theta=\frac{2k^{3/2}\left(\sinh(2\sqrt{k})+
\sin(2\sqrt{k})\right)}{k(\sinh(2\sqrt{k})+\sin(2\sqrt{k}))-(\cosh(2\sqrt{k})-\cos(2\sqrt{k}))}\end{equation}
Optimization over the sensitivity leads to an enhancement of
$2/3$ in comparison with the steady state case. An even bigger
gain would have been obtained if a projective measurement at the
end of the continuous observation were allowed. A realistic way
to implement this would be to make the sensitivity very large at
the end of the measurement.  If the matrix $P(1)$ is the solution
of the Riccati equation (\ref{Ric}) at time $t=1$, some
straightforward calculations show that a projective position
measurement reduces the estimator variance by $P_{(3,1)}^2/P_{(1,1)}$.
The optimal initial conditions are still given by ($\langle\Delta
p^2\rangle\ll 1$) and $\langle\Delta x^2\rangle\gg 1$. The exact
expression of the variance on the estimate in function of the
sensitivity $k$ is then given by:
\begin{equation}\sigma_\theta=\frac{4k^{3/4}(\cosh(2\sqrt{k})+
\cos(2\sqrt{k}))}{k(\cosh(2\sqrt{k})+\cos(2\sqrt{k}))-(\sinh(2\sqrt{k})+\sin(2\sqrt{k}))}\end{equation}
Minimization over the sensitivity leads to the equation:
\begin{equation}\sigma_\theta\simeq 0.752 \frac{4\hbar m}{t^3}\end{equation}
Therefore we have modestly beaten the usual standard quantum limit
by optimally preparing the Gaussian wavepacket and doing a von
Neumann measurement at the end of the continuous measurement. This
shows that a continuous measurement together with a projective
measurement at the end on a optimally prepared state can reveal
more information than only projective measurements. In other
words, the balance information gain versus disturbance is a
little bit in favor of continuous measurement. Although noise is
continuously fed into the system by the sensor, we can extract
more information about the classical force.

An even better performance can be obtained if we vary the sensitivity
continuously during the measurement ({\em sensitivity scheduling}). It is
indeed the case that backaction noise introduced in the beginning of the
measurement does more harm than backaction noise at the end of the
measurement, as the random momentum kicks delivered at any given time corrupt
all subsequent position readouts. In terms of systems theory, the optimal
sensitivity as a function of time can be found by solving an optimal control
problem associated with equation (\ref{Ric}). This optimal control can be
determined by solving a Bellman equation by using techniques of dynamic
programming~\cite{maybeck1982a}. Due to the nonlinearity of the
Riccati equation however, this 
cannot be done analytically. The optimal sensitivity at time $\tau$ however
can easily be obtained: it tends to a delta-Dirac impulse such as to mimic a
projective position measurement, reducing the variance with
$P_{(3,1)}^2/P_{(1,1)}$. Defining the cost-function
$K=P_{(3,3)}(\tau)-P_{(3,1)}^2(\tau)/P_{(1,1)}(\tau)$, the optimal control
problem is then well defined and can be solved numerically. Therefore we
discretize the total time in for example 50 intervals, and in each interval
we assume the sensitivity has a  constant value $k_j$. The solution can then
found by applying some kind of steepest descent algorithm over these 50
variables $\{k_j\}$. It turns out that the optimal $k(t)$ in the case of a
free particle ($\omega=0$) is a smooth monotonously but slowly increasing
function of time. In this free particle case, the optimal time-varying
sensitivity only leads to a marginal gain: the numerical optimization shows
that the variance of the estimate becomes exactly equal to a factor $3/4$ of
the usual Standard Quantum Limit (\ref{SQLclas}). Nevertheless, we can
present this result as a generalization of the usual SQL to include
strategies with sensitivity scheduling:
\begin{equation}\sigma_\theta= \frac{3\hbar m}{t^3}\end{equation}

Much greater improvements can be expected from the application of sensitivity
scheduling to the case of a continuously observed harmonic oscillator.
Indeed, the variance on the position of such a particle is small in the
middle of the well and at the borders, while it is big elsewhere. Therefore
the sensitivity should be varied in a sinusoidal manner, such as to measure
more precisely at the positions where the variance is small. The optimal
variation of sensitivity in time could be determined by solving a similar
optimal control problem to the one explained in the previous paragraph. In
the limit where projective measurements are allowed, one expects that the
optimal variation of sensitivity should correspond to stroboscopic
measurement \cite{Braginsky}, which is indeed well-known to beat the usual
standard quantum limit.

\subsection{Detection of non-stationary signals}
The techniques introduced in the previous sections can also be used for the
estimation of non-stationary signals, as one would have for example in the
problem of gravitational wave detection when the arrival time of the signal
is unknown. Suppose for example that we know that the signal to detect is of
the form $\theta(t-t_1)=\theta_0 f(t-t_1)$ with $f(\tau)$ known but amplitude
$\theta_0$ and arrival time $t_1$ unknown. An effective non-stationary
measurement strategy can in fact be implemented by constructing a Kalman
filter for system (\ref{cse}) assuming that $\theta=0$ (assuming $f(\tau)=0$
for $\tau<0$). At times $t<t_1$, the quantity $d\xi-\bar{x}(t)dt$ is by
construction white noise with variance $dt/2k(t)$. From time $t\geq t_1$ on
however, the force will bias this white noise by an amount
$\int_{t_1}^{t}dt'g(t,t')\theta(t')$ as the $\theta=0$ Kalman filter models
the wrong system. This bias will be detectable once it transcends the white
noise at time $t_1+\Delta t$:
\begin{equation}\int_{t_1}^{t_1+\Delta t}dt\int_{t_1}^{t}dt'g(t,t')\theta_0
f(t'-t_1)\geq\sqrt{\int_{t_1}^{t_1+\Delta
t}\frac{dt}{2k(t)}}\label{transcent}\end{equation} The goal is
now to make this $\Delta t$ as small as possible. The previous
equation can again be solved analytically if one has a constant
sensitivity and steady state conditions. To make things easier we
assume that the observed particle is free ($\omega=0$), although
all calculations can be performed in the more general case too.
Let us first assume that the signal to detect is a kick at time
$t_1$: $f(t-t_1)\simeq\delta(t-t_1)\tau$ with $\tau$ some measure
of the duration of the kick \cite{Braginsky}. Introducing the
dimensionless parameter $\kappa=\Delta t\sqrt{\hbar k/2m}$, the
previous inequality becomes:
\begin{eqnarray}\theta_0&\geq&\frac{1}{\tau}\sqrt{\frac{\hbar m}{\Delta
t}}\frac{\kappa}{1-\exp(-\kappa)(\cos(\kappa)+\sin(\kappa))}\\
&\geq&\frac{2}{\tau}\sqrt{\frac{\hbar m}{\Delta t}}\end{eqnarray}
In the last step the optimal $\kappa$, related to the optimal
sensitivity $k$, was chosen. The meaning of this equation is
clear: a kick with an amplitude $\theta_0$ will only be observed
after a time span $\Delta t=4\hbar m/\tau^2\theta_0^2$. Moreover,
the sensitivity has to scale inversely with the square root of
$\Delta t$.

An analogous treatment applies to the case of a constant force
$f(t-t_1)=u_{[0,\infty]}(t-t_1)$. In this case inequality (\ref{transcent})
becomes: \begin{eqnarray}\theta_0&\geq& \sqrt{\frac{\hbar m}{\Delta
t^3}}\frac{\kappa^2}{\exp(-\kappa)\cos(\kappa)+\kappa-1}\\
&\geq&4.25\sqrt{\frac{\hbar m}{\Delta t^3}}\end{eqnarray} As expected, we
recover the well known standard quantum limit, but now in a different set-up.

The previous arguments can be refined by using techniques of classical
detection theory such as the concept of the matched filter. The results will
however be qualitatively similar to the previous ones.

More advanced detection schemes can also be constructed by {\em adaptively}
changing the sensitivity as a real-time function of the measurement record
\cite{wiseman1995}. A possible application of this is a scheme for the
detection of a signal with unknown arrival time: first one chooses the
optimal sensitivity for estimating the arrival time, and from the moment on
the signal is detected the sensitivity is brought to its optimal value for
detecting the amplitude of the signal. More sophisticated versions of this
adaptive measurement could be very useful in realistic stroboscopic
measurements where the initial phase of the harmonic oscillator is unknown,
as the measurement sensitivity could be made a real-time function of the
estimated particle position.

\begin{acknowledgements}
This work was supported by the National Science Foundation, the
A.\ P.\ Sloan Foundation and the K.\ U.\ Leuven. FV is grateful to Bart
De Moor.
\end{acknowledgements}

\end{document}